\documentclass[10pt]{article}
\usepackage{cite}
\usepackage{epsfig}
\usepackage[dvips,usenames]{color}
\usepackage{color}
\usepackage{rotating}
\usepackage{amsmath,amssymb,amsfonts}
\usepackage{latexsym}
\begin{document}
\setcounter{section}{0}
\setcounter{equation}{0}
\setcounter{figure}{0}
\setcounter{table}{0}
\setcounter{footnote}{0}
\begin{center}
{\bf\Large The $1+1$ Dimensional Abelian Higgs Model Revisited:}

\vspace{5pt}

{\bf\Large Physical Sector and Solitons}\footnote{Contribution to the Proceedings of the Fifth International
Workshop on Contemporary Problems in Mathematical Physics, Cotonou, Republic of Benin, October 27--November 2, 2007,
eds. Jan Govaerts and M. Norbert Hounkonnou (International Chair in Mathematical Physics and Applications,
ICMPA-UNESCO, Cotonou, Republic of Benin, 2008), pp.~164--169.}
\end{center}
\vspace{10pt}
\begin{center}
Laure GOUBA$^{\dagger,\ddagger}$, Jan GOVAERTS$^{*,\star,\diamondsuit}$
and M. Norbert HOUNKONNOU$^{\diamondsuit}$\\
\vspace{5pt}
$^\dagger${\sl National Institute for Theoretical Physics (NITheP),\\
Stellenbosch Institute for Advanced Study (STIAS),\\
Private Bag X1, Matieland 7602, Republic of South Africa}\\
{\it E-Mail: gouba@sun.ac.za}\\
\vspace{7pt}
$^\ddagger${\sl African Institute for Mathematical Sciences (AIMS),\\
6 Melrose Road, 7945 Muizenberg, Republic of South Africa}\\
{\it E-Mail: laure@aims.ac.za}\\
\vspace{7pt}
$^{*}${\sl Center for Particle Physics and Phenomenology (CP3),\\
Institut de Physique Nucl\'eaire, Universit\'e catholique de Louvain (U.C.L.),\\
2, Chemin du Cyclotron, B-1348 Louvain-la-Neuve, Belgium}\\
{\it E-Mail: Jan.Govaerts@uclouvain.be}\\
\vspace{7pt}
$^\star${\sl Fellow, Stellenbosch Institute for Advanced Study (STIAS),\\
7600 Stellenbosch, Republic of South Africa}\\
\vspace{7pt}
$^{\diamondsuit}${\sl International Chair in Mathematical Physics and Applications (ICMPA-UNESCO Chair),\\
University of Abomey--Calavi, 072 B. P. 50, Cotonou, Republic of Benin}\\
{\it E-Mail: norbert.hounkonnou@cipma.uac.bj}
\end{center}

\vspace{15pt}

\begin{quote}
In this paper the two dimensional abelian Higgs model
is revisited. We show that in the physical sector, the solutions
to the Euler--Lagrange equations include solitons.
\end{quote}

\vspace{10pt}

\section{Introduction}
\label{Sec1.Gouba2}

The two dimensional abelian Higgs model is revisited. Using the Dirac
formalism for constrained systems, it has been established \cite{Lau} 
that this model, in its gauge invariant physical sector, corresponds to the
coupling of a pseudoscalar field, namely the electric field,
with a real scalar field, as a matter of fact the Higgs field which is the radial
component of the original complex scalar field in field configuration space. At the classical level,
the Euler--Lagrange equations of motion lead to a system of coupled non-linear equations
of which the spectrum of solutions includes solitons. The linearised spectrum of 
fluctuations of these solitonic solutions has been identified in order to ascertain
their classical stability.

This contribution is organised as follows. Section \ref{Sec2.Gouba2} discusses
the Euler--Lagrange equations of motion. Upon compactification of the spatial dimension into
a circle, in Section \ref{Sec3.Gouba2} static solutions to these equations are constructed  in closed
analytic form in terms of the Jacobi elliptic functions. In Section \ref{Sec4.Gouba2} the linearised
spectrum of fluctuations for these classical solutions is identified. Some concluding remarks
are provided in Section \ref{Sec5.Gouba2}.

\section{The Euler--Lagrange Equations of Motion}
\label{Sec2.Gouba2}

The $1+1$ dimensional abelian Higgs model is described by the 
Lagrangian density
\begin{eqnarray}\nonumber
{\mathcal{L}} =-\frac{1}{4}\left[\partial_\mu A_\nu -\partial_\nu A_\mu \right]
\left[\partial^\mu A^\nu -\partial^\nu A^\mu\right] +
\vert\left(\partial_\mu + ie A_\mu\right)\phi \vert^2 - V(|\phi|),
\end{eqnarray}
where $A_\mu$ is the gauge field with gauge coupling constant $e$ and $\phi$ a complex scalar field
with self-interactions described by the U(1) gauge invariant potential $V(|\phi|)$.
Choosing  a parametrisation of the complex field as $\phi = \rho e^{i\varphi}/\sqrt{2}$ and 
through the Dirac formalism for constrained systems \cite{Gov} the model in its physical
sector is described by \cite{Lau}
\begin{equation}
{\mathcal{L}}_{\textrm{phys}} = \frac{1}{2}\frac{1}{\rho^2}(\partial_\mu B)^2 
-\frac{1}{2} e^2 B^2 +\frac{1}{2}(\partial_\mu\rho)^2 -V(\rho)
- \partial_0[\frac{1}{\rho^2}B\partial_0 B] +\partial_1\left[B(\partial_0\varphi +e A_0)\right],
\label{sur2}
\end{equation}
where $B = -\frac{1}{e}E$ and $E$ is the electric field.
The Euler--Lagrange equations of motion for $\rho$ and $B$ are, respectively,
\begin{equation}
\partial_\mu^2\rho + \frac{1}{\rho^3}(\partial_\mu B)^2 +
\frac{\partial V}{\partial\rho} =0, \qquad
\partial_\mu\left(\frac{1}{\rho^2}\partial^\mu B\right) + e^2 B =0.
\label{420}
\end{equation}
These equations are non-linear and coupled, and not all their
solutions can be analytically found. However it might be possible
to find analytical solutions for static states of finite energy.
In order to reduce the difficulties of solving these equations, we consider
configurations where the electric field vanishes, $E(t,x)=0$. In fact
the total energy is $E_{\rm phys} = E_k + E_p$, where $E_k$ is the total kinetic energy
and $E_p$ the total potential energy of the fields, with,
\begin{eqnarray}\nonumber
E_k =\int_{-L}^L dx\left(\frac{1}{2\rho^2}(\frac{\partial}{\partial t} B)^2 
+\frac{1}{2}(\frac{\partial}{\partial t}\rho)^2 \right),
\end{eqnarray}
and
\begin{eqnarray}\nonumber
E_p =\int_{-L}^L dx\left(\frac{1}{2}\frac{1}{\rho^2}(\frac{\partial}{\partial x}B)^2 + \frac{1}{2}e^2 B^2
 +\frac{1}{2}(\frac{\partial}{\partial x}\rho)^2 
+V(\rho)\right).
\end{eqnarray}
{}From these expressions, one notices that if $B$ is non zero but $\rho$
vanishes for some value of $x$, we have a singularity in the equations of motion
and furthermore the energy may become infinite, unless the electric field vanishes
faster than $\rho$ in such a manner that the quotient of these zeros remains finite.
But such behaviour induces large spatial gradients in both $\rho$ and $B$, implying a large
value for the total energy. Hence in order to minimise the energy, one needs to consider
configurations with $B=0$. Another advantage of this restriction is that
the equations are no longer coupled and we simply have to solve
\begin{equation}\label{421}
B =0,\qquad
\partial_\mu^2\rho +\frac{\partial V}{\partial\rho}  =0.
\end{equation}
A further assumption is required though, since it is still difficult
to construct all the time dependent solutions to the above equation (except for
those obtained by Lorentz boosts from a static solution). Moreover, any time
dependence for a solution increases its total energy. Consequently
we restrict further to static configurations in which $\rho$ only depends
on the space variable $x$, leading to the single non trivial equation,
\begin{equation}\label{422}
B(t,x) = 0,\qquad
\frac{d^2\rho(x)}{dx^2} -\frac{d V(\rho(x))}{d\rho(x)} = 0,
\end{equation}
where for the potential energy henceforth the Higgs choice will be made,
$V(\rho) = (1/8)M^2\left(\rho^2-\rho_0^2\right)^2$, with $M>0$ a mass scale and
$\rho_0$ the expectation value of the scalar field.

\section{Static Solutions}
\label{Sec3.Gouba2}

In order to solve the above equation, let us choose to compactify space
into a circle of length $2L$ with $-L\le x\le L$.
For any constant values of $\rho$ associated to the minima of
the potential, where the potential also vanishes given our choice of subtraction
constant, the non-linear equation in (\ref{421})  is satisfied.
These constant solutions $B(t,x) =0$, $\rho(t,x) =\pm\rho_0$ correspond to vacuum configurations,
of which the total energy values are minimal and vanish.

\pagebreak

Besides these configurations, we also have non constant static solutions which
behave like solitons. Given the equation for $\rho$,
\begin{eqnarray}\label{423}
\rho'' -\frac{\partial V}{\partial\rho} = 0,
\end{eqnarray}
where $\rho''(x)$ stands for the double derivative with respect to $x$, there exists
a conservation law expressed as
\begin{eqnarray}\label{juil}
\frac{1}{2}((\rho'(x))^2 = V(\rho) + C_0,
\end{eqnarray}
$C_0$ being some integration constant. In this form the solution is readily
obtained by quadrature. On the circle of length $2L$ one finds,
\begin{eqnarray}\nonumber
\rho_s(x) =\pm\rho_0\sqrt{\frac{2k^2}{1+k^2}}\ {\rm sn} 
\left[\frac{1}{2}M\rho_0\sqrt{\frac{2}{1+k^2}} x\right].
\end{eqnarray}
it being understood that the solutions obey either periodic or anti-periodic boundary conditions
given the remaining freedom in the sign of $\rho$ existing for the choice of polar parametrisation
of the complex scalar field $\phi$ \cite{Lau}. These solutions are thus expressed in terms of the
Jacobi elliptic functions \cite{Arsc}. According to whether a periodic or an anti-periodic
solution is obtained, the remaining single integration constant is related to the elliptic modulus $k$
through the condition,
\begin{eqnarray}\nonumber
LM\rho_0 &=& 4(n+r)K(k)\sqrt{\frac{1+k^2}{2}},\qquad n\in\mathbb{N},\\
\nonumber
 r &=& \left\{
\begin{array}{cc}
0 : &\:\:\textrm{periodic},\\
1/2: & \:\:\textrm{antiperiodic}.
\end{array}\right. 
\end{eqnarray}
Their total energy is represented through the expression
\begin{eqnarray}\nonumber
E_{\rm phys}=\int_{-\infty}^{+\infty} dx\left\{\frac{1}{4}M^2
\left(\rho^2_s -\rho_0^2\right)^2 +C_0\right\}.
\end{eqnarray}
By substituting the above explicit expression, one finds
\begin{equation}
\nonumber
E_{\rm phys}=\frac{LM^2\rho_0^4}{2(n+r)K(k)}\int_0^{2(n+r)K(k)} dy
\left\{ \left[\frac{1}{4} +\frac{k^2}{(1+k^2)^2}\right]
- \frac{2k^2}{(1+k^2)}\ {\rm sn}^2y +\frac{2k^4}{(1+k^2)^2}\ {\rm sn}^4y\right\}.
\end{equation}
This quantity is finite since $ 0\le k^2\le 1$ et $ 0\le\ {\rm sn}^2y \le 1$. This very fact together with
the spatial profile $\rho(x)$ of these configurations justifies their interpretation as solitons.
In particular the periodic solution with $n=1$ in the decompactification limit $L\rightarrow\infty$
reduces to the celebrated kink solutions of the $\phi^4$ real scalar field theory in 1+1 dimensions.

\section{Spectrum of Fluctuations}
\label{Sec4.Gouba2}

Given the explicit solutions of the previous Section, in the present
Section we address the issue of the classical stability under linearised
fluctuations in field configuration space. This requires the
computation of the spectrum of fluctuation eigenvalues, to ascertain
that none of these is negative, which otherwise would establish that some
modes have an unbounded above exponentially growing amplitude, spelling disaster for the
corresponding solution.

Let us consider arbitrary time- and space-dependent fluctuations around
the identified solutions, $B(t,x) =\delta B(t,x)$, $\rho(t,x) =\rho_s(x)+\delta\rho(t,x)$.
The corresponding linearised Lagrangian density, expanded to second order in these fluctuations, is
\begin{eqnarray}\nonumber
\begin{array}{r l}
{\mathcal L}=&\frac{1}{2\rho^2_s}\left(\partial_\mu\delta B\right)
\left(\partial^\mu\delta B\right)-\frac{1}{2}e^2\left(\delta B\right)^2
+\frac{1}{2}\left(\partial_\mu \rho_s\right)
\left(\partial^\mu\rho_s\right)+\left(\partial_\mu\rho_s\right)
\left(\partial^\mu\delta\rho\right)\\ 
&\\\nonumber
+&\frac{1}{2}
\left(\partial_\mu\delta\rho\right)\left(\partial^\mu\delta\rho\right)
-V(\rho_s)-\delta\rho V'(\rho_s)-\frac{1}{2}\left(\delta\rho\right)^2
V''(\rho_s)\ .
\end{array}
\end{eqnarray}
Applying the variational principle in the fields $\delta B$ and $\delta\rho$,
the linearised Euler--Lagrange equations for these fields read
\begin{equation}
\label{555}
\partial_\mu\partial^\mu\delta\rho+V''(\rho_s)\delta\rho = 0,\qquad
\partial_\mu\left[\frac{1}{\rho^2_s}\partial^\mu\delta B\right]+e^2\delta B = 0.
\end{equation}
These equations being linear in the field variables, a normal mode expansion is warranted,
with a spectrum of eigenfrequencies to be identified through the following {\it ansatz\/}
for time dependence (the general solution then being constructed through linear
combinations),
\begin{equation}\label{ffe}
\delta\rho(t,x) = f(x)e^{-i\omega_1t} + f^{*}(x)e^{i\omega_1t},\qquad
\delta B(x,t)   = g(x)e^{-i\omega_2t} + g^{*}(x)e^{i\omega_2t}.
\end{equation}
If there exist solutions for which either of the eigenfrequencies $\omega_1$
or $\omega_2$ is pure imaginary, namely such that $\omega^2_i<0$ ($i=1,2$),
this would imply an exponential run-away time dependence for at least one of the linearised
fluctuation modes, hence instability of the corresponding classical solution.
In the above parametrisation of the normal modes, $f(x)$ and $g(x)$ stand
for complex-valued functions on the circle, with $g(x)$ periodic and $f(x)$
periodic or anti-periodic according to the periodicity properties of the classical
solution of which the stability needs to be established.

By direct substitution of the above {\it ansatz\/} into (\ref{ffe}),
the following eigenvalue equations are derived,
\begin{eqnarray}\label{557}
\left(-\frac{d^2}{dx^2}+V''(\rho_s)\right)f(x)=\omega^2_1\,f(x)\ ,\\\label{600}
\left[-\rho^2_s(x)\frac{d}{dx}\frac{1}{\rho^2_s(x)}\frac{d}{dx}
+e^2\rho^2_s(x)\right]g(x)=\omega^2_2\,g(x)\ .
\end{eqnarray}
In the case of the vacuum configuration, $B=0$, $\rho=\pm\rho_0$, 
the solutions to (\ref{557}) and (\ref{600}) are readily
constructed through a Fourier series analysis on the circle for
the unknown functions $g$ and $f$, with
$g(x) = 1/(2L)\sum_{n=-\infty}^{+\infty}e^{i\frac{\pi n}{L}x}g_n$ and
$f(x) = 1/(2L)\sum_{n=-\infty}^{+\infty}e^{i\frac{\pi n}{L}x}g_n$. 
The spectrum is then found to be given as
\begin{eqnarray} \nonumber
f_n\ :\qquad \omega^2_1 &=& \left(\frac{\pi n}{L}\right)^2+M^2\rho^2_0,\\\nonumber
g_n\ :\qquad \omega^2_2 &=& \left(\frac{\pi n}{L}\right)^2+e^2\rho^2_0,
\end{eqnarray}
which is indeed positive definite. As it should, the vacuum configuration is indeed
stable against all possible fluctuations in the fields.

For the non-trivial soliton configurations, the eigenvalue problem reads,
\begin{eqnarray}\label{lame}
\left[\frac{d^2}{dy^2} -6k^2\ {\rm sn}^2y\right] f(y) =
-\left[\frac{2\omega^2_1}{M^2\rho_0^2}+1\right](1+k^2) f(y),
\end{eqnarray}
\begin{eqnarray}\label{schrod}
\left[-\frac{d^2}{dy^2} + W_B(y)\right]\psi(y)
=\frac{2(1+k^2)\omega_2^2}{M^2\rho_0^2}\psi(y),
\end{eqnarray}
where 
\begin{eqnarray}\label{potent}
W_B(y) = 2\left[\frac{2e^2k^2}{M^2}\ {\rm sn}^2y 
-\frac{1+k^2}{2} +\frac{1}{{\rm sn}^2y}\right].
\end{eqnarray}
In order to obtain (\ref{schrod}), the following change of variable was introduced in (\ref{600}),
$ g(x) = \rho_s(x)\psi(x)$. As a matter of fact, (\ref{lame}) is a Lam\'e equation,
of which the solutions have been discussed and classified in Ref.\cite{Arsc}, with the
results listed in the Table below.

\begin{table}[t]
\centering
\begin{tabular}{|c|c|}
\hline
 & \\
$f(y)$ & $ \omega_1^2$\\[5pt]
\hline\hline
 & \\
${\rm sn}\,y\cdot{\rm cn}\,y$ & $\frac{1}{2}M^2\rho_0^2\frac{3}{k^2+1}$\\[5pt]
\hline
 & \\
${\rm sn}\,y\cdot{\rm dn}\,y$ & $\frac{1}{2}M^2\rho_0^2\frac{3k^2}{1+k^2}$\\[5pt]
\hline
 & \\
${\rm cn}\,y\cdot{\rm dn}\,y$ & $0$ \\[5pt]
\hline
 & \\
${\rm sn}^2 y - \frac{1}{3k^2}(1+k^2\pm\sqrt{1-k^2(1-k^2)}$ & $\frac{1}{2}M^2\rho_0^2\left(1\mp 2\frac{\sqrt{1-k^2(1-k^2)}}{1+k^2}\right)$\\[5pt]
\hline
\end{tabular} 
\end{table}

As may be seen from those results, the eigenfunctions $f(y) = {\rm sn}\,y\cdot{\rm cn}\,y$
and $f(y) ={\rm sn}\,y\cdot{\rm dn}\,y$ both have a positive eigenspectrum of $\omega^2_1$
values and are both antiperiodic. However the last three solutions in the Table correspond
to periodic functions, one of which is the zero mode associated to infinitesimal spatial
translations, while among the other two there always exists one with a strictly negative
eigenvalue $\omega^2_1$. Consequently the solitons possessing a periodic boundary condition
on the circle are unstable against fluctuations corresponding to that specific mode.

For what concerns the function $g(x)$ and its fluctuation spectral equation (\ref{schrod}),
the latter eigenvalue problem being equivalent to solving the Schr\"odinger equation for
the potential $W_B(y)$, the issue may also be addressed from that point of view. Even though
the Schr\"odinger equation may not be explicitly solve for that choice of potential, using the
fact that the potential (\ref{potent}) is positive, implies in any case that the spectrum 
of $\omega^2_2$ eigenvalues is likewise positive. No instability may arise in the $g(x)$,
namely the $B$ sector of fluctuations.

\section{Concluding Remarks}
\label{Sec5.Gouba2}

The vacuum configuration has thus been confirmed to be stable against all
fluctuations. Non-trivial static soliton configurations are also stable,
but only in the sector of anti-periodic solitons and thus an anti-periodic
boundary condition on the function $f(x)$. Periodic solitons, though,
are always unstable.

Even though this work has some overlap with some previous studies, its originality lies
with the fact that it considers specifically only the physical sector of the 1+1 dimensional
abelian U(1) Higgs model without applying any gauge fixing procedure whatsoever.
In particular the factorisation between the actual physical sector and the decoupled
gauge variant sector offers two advantages. First, that no artefacts due to gauge fixing
are introduced. Second, that potential spurious instabilities lying solely within the
gauge variant and unphysical sector are avoided from the outset, a feature any other approach
having been developed so far cannot achieve. All these approaches have until now relied
on some gauge fixing procedure rather than applying
such a gauge invariant and physical factorisation. Hence these approaches runs the risk of
identifying instabilities which in fact are pure gauge and thus unphysical. Such difficulties
are avoided from the outset within our approach.

When the potential $V(\rho)$ is taken to be the Higgs potential, as we did,
the static solutions in $\rho$ with $B=0$ are in fact those of Refs.\cite{Mant,Bri}.
These authors also solve the model of a single complex scalar field on the circle,
while in Ref.\cite{Briha}, the authors consider the $1+1$ abelian Higgs Model on the circle,
but in contradistinction to what we have done, they apply a gauge fixing procedure
in terms of the gauge field components $A_0$ and $A_1$.
We recover the same solutions on the circle. Our work has established
that the soliton configurations with anti-periodic boundary conditions
are stable. In Refs.\cite{Bri, Briha}, the static soliton solutions with
periodic boundary conditions are also found to be unstable.

A possible continuation of the present study would be the computation of the
complete spectrum of fluctuation eigenfrequencies for the stable and unstable
classical soliton configurations, inclusive of the electromagnetic sector
contributions but retaining only the physical degrees of freedom,
in order to  determine the quantum corrections of the mass spectrum of these solitonic
field configurations of the 1+1 dimensional abelian U(1) Higgs model, whether on the
circle or on the real line.

\section*{Acknowledgements}

This work, part of Laure Gouba's Ph.D. thesis, was revisited
during her stay at the African Institute for Mathematical Sciences (AIMS)
as a Postdoctoral Fellow and Teaching Assistant. L.G. would like to
thank Profs. Neil Turok, founder of AIMS, and Fritz Hahne, Director
of AIMS, as well as the AIMS family for their support and hospitality.

J.G. is grateful to Profs. Hendrik Geyer, Bernard Lategan and Frederik Scholtz for the support and the hospitality
of the Stellenbosch Institute for Advanced Study (STIAS) with the grant of a Special STIAS Fellowship
which made a stay at STIAS and NITheP possible in April-May 2008. He acknowledges the Abdus Salam International
Centre for Theoretical Physics (ICTP, Trieste, Italy) Visiting Scholar Programme in support of
a Visiting Professorship at the ICMPA-UNESCO (Republic of Benin).
J.G.'s work is also supported by the Institut Interuniversitaire des Sciences Nucl\'eaires, and by
the Belgian Federal Office for Scientific, Technical and Cultural Affairs through
the Interuniversity Attraction Poles (IAP) P6/11.

\end{document}